\documentclass[reprint, amsmath,amssymb, aps, nofootinbib]{revtex4-1}

\usepackage{natbib}
\usepackage{graphicx}
\usepackage{dcolumn}
\usepackage{bm}

\makeatletter
\newcommand*{\rom}[1]{\expandafter\@slowromancap\romannumeral #1@}
\makeatother

\begin{document}

\title{Characteristic times  \\ in the standard map}

\author{Mirella Harsoula}
 \email{mharsoul@academyofathens.gr} 
\author{Kostas Karamanos}%
 \email{koskaraman@gmail.com}
 \author{George Contopoulos}
 \email{gcontop@academyofathens.gr}
\affiliation{%
 Research Center for Astronomy,
           Academy of Athens\\ Soranou Efesiou 4, GR-115 27 Athens, Greece}
\date{\today}

\begin{abstract}

We study and compare three characteristic times of the standard map, the Lyapunov time $t_L$, the Poincar$\acute{e}$ recurrence time $t_r$ and the stickiness (or escape) time $t_{st}$. The Lyapunov time is the inverse of the Lyapunov characteristic number ($LCN$) and in general is quite small. We find empirical relations for the $LCN$ as a function of the nonlinearity parameter $K$ and of the chaotic area $A$. We also find empirical relations for the Poincar$\acute{e}$ recurrence time $t_r$ as a function of the nonlinearity parameter $ K $, of the chaotic area $A$ and of the size of the box of initial conditions $\epsilon$. As a consequence we find relations between $t_r$ and $LCN$. We compare the distributions of the stickiness time and the Poincar$\acute{e}$ recurrence time. The stickiness time inside the sticky regions at the boundary of the islands of stability is orders of magnitude smaller than the  Poincar$\acute{e}$ recurrence time $t_r$ and this affects the  diffusion exponent $\mu$, which converges always to the value $\mu=1$. This is shown in an extreme stickiness case. The diffusion is anomalous (ballistic motion) inside the accelerator mode islands of stability with $\mu=2$ but it is normal everywhere outside the islands with $\mu=1$.  In a particular case of extreme stickiness we find the hierarchy of islands around islands.  

\begin{description}
\item[PACS numbers]
05.45.-a
\end{description}
\end{abstract}

\maketitle

\section{Introduction}

Chaos in Hamiltonian systems and 2-D mappings 
 is in general incomplete and nonergodic due to the existence of islands of stability with KAM
invariant tori inside them. The set of islands is fractal, and the region close to and outside the
islands' boundary is sticky and plays a
crucial role in the system's kinetics. 

The combination of the ``normal'' random walk of trajectories inside the large chaotic sea far away from the islands of stability with flights and trappings of ``sticky'' trajectories near the boundaries of the islands, creates a new kind of kinetics, called
``strange kinetics''.
The term ``strange kinetics'' was introduced in \cite{1971sfmm.book.....M}  to describe a class of processes that  lead to anomalous transport. 
In order to study such complicated phase spaces it is important to calculate the characteristic times of the mappings or Hamiltonian systems, i.e. the Lyapunov time, the Poincar$\acute{e}$ recurrence time and the stickiness time, and find the relations between each other. 
These characteristic times give important information about the complexity of the system
and they can be related to other interesting measures like the fractal (or Hausdorff) dimension and the diffusion exponent.

Another important issue is the dependence of these characteristic times on the initial conditions.
If we choose, for example, a set of initial conditions situated in the extreme sticky region around islands of stability,
the orbits will present the same behavior as an orbit inside the island  for long enough time (depending on the distance from the last KAM curve) but later on they will behave as chaotic orbits manifesting normal diffusion.  
Thus these characteristic times can help one to study the
complexity of the orbital structure.

In order to make our study we use a well known 2-D mapping, namely the standard (or Chirikov) map:

\begin{eqnarray}\label{stand}
\nonumber & x' = x + y' \quad (\mathrm{mod}\,1)\\&
y' = y +\frac{K}{2 \pi}  \sin (2 \pi x ) \quad (\mathrm{mod}\,1)
\end{eqnarray}

This is a typical simple example of a 2-D map that has been studied extensively up to now. However, there are still certain questions that have not been clarified completely. These refer mainly to the characteristic times of the standard map, namely the Lyapunov time, the Poincar$\acute{e}$ recurrence time and the stickiness time.
In the present paper we discuss these times and their relations.

In the literature we find some interesting applications of the characteristic times in mappings and Hamiltonians.
For example, in Celestial mechanics it has been shown that there is a power-law dependence of the recurrence times on Lyapunov times and a power-law decay in the tails of the recurrence distributions. Some characteristic examples are given for the case of the three body problem \cite{2010PhRvE..81f6216S} or for the case of asteroids' orbits \cite{1993ApJ...406L..35L}. Moreover, studies in dielectric cavities show different behavior of the survival probability time distributions in integrable and in mixed or chaotic cases (see \cite{2006PhRvE..73c6207R}). The laws given for these quantities agree with the laws given for the distribution of the Poincar$\acute{e}$  recurrence times in the case of the 2-D mapping described in the present article.  

The paper is organised as follows: In section \rom{2} we  give analytical relations for the Lyapunov characteristic number as a function of the perturbation parameter $K$ and of the available chaotic area  $A$. Moreover, we discuss its dependence on the initial conditions. In section \rom{3} we study the behavior and the distribution of the Poincar$\acute{e}$ recurrence times below and above the critical value $K_c$ as well as the law giving its relation with the Lyapunov characteristic number and the corresponding available chaotic area in the phase space. We also connect the mean Poincar$\acute{e}$ recurrence time with the fractal dimension $d_w$ of the mapping. In section \rom{4} we discuss the stickiness (or escape) time. In section \rom{5} we study the dependence of the type of diffusion (normal or anomalous)  on the initial conditions of the orbit and the role of stickiness. In section \rom{6} we make a comparison between the characteristic times in extreme stickiness cases.  In section \rom{7} we analyze the case of self-similarity of islands and finally in section \rom{8} we summarize our conclusions.

\section{The Lyapunov time}
 In \cite{1994JPhA...27.4899V} the authors give the definition of the stretching number $\alpha$ and argue that the spectrum of this number for a specific nonlinearity parameter $K$ is invariant for the case of the standard map. The stretching number is defined as follows:
 
 \begin{equation}
 \alpha =\ln \Big(\frac{ds'}{ds}\Big)
 \end{equation} 
 where $ds=\sqrt{dx^2+dy^2}$ is an infinitesimal initial separation between two orbits and $ds'=\sqrt{dx'^2+dy'^2}$ is the separation after one iteration of the map. In general, for a two dimensional area preserving mapping:
 
 \begin{equation}
x' = f(x,y,K)   ~~~y'=g(x,y,K) 
 \end{equation} 
 where $K$ is the nonlinearity parameter,
 the stretching number takes the form:
  \begin{equation}\label{alfa}
\alpha=\frac{1}{2} \ln \Bigg(\frac{(\frac{\partial f}{\partial x}+\frac{\partial f}{\partial y} y_x)^2+(\frac{\partial g}{\partial x}+\frac{\partial g}{\partial y} y_x)^2}{1+y_x^2}\Bigg)
 \end{equation} 
 where $y_x=dy/dx$.
 In the case of the standard map eq. (\ref{alfa}) becomes:
  \begin{equation}\label{alfan}
\alpha=\frac{1}{2} \ln \Bigg(\frac{[y_x+K \cos(2 \pi x)]^2+[y_x+(1+K \cos(2 \pi x))]^2}{2}\Bigg)
 \end{equation} 
 
The finite time Lyapunov characteristic number ($ftLCN$) in the chaotic region is defined by the relation: $ftLCN=(1/t) \ln \Big(ds(t)/ds(0)\Big)$, where $ds(0)$ is an infinitesimal initial separation between two orbits and $ds(t)$ is their separation at time $t$.  The $ftLCN$ can be obtained also from the mean value of the stretching number $\alpha$ over a time span $t$:

\begin{equation}\label{lcn}
ftLCN(t)=\langle\alpha\rangle=\frac{\sum_{i=1}^{t} \alpha_i} {t} 
\end{equation}
The Lyapunov characteristic number $LCN$ is the limit of $ftLCN$ for $t \rightarrow \infty$. We must point out here that the time $t$ in the case of a 2-D mapping is given by the number of iterations.
By integrating chaotic orbits for long enough time to secure convergence we can finally extract the $LCN$ as a function of: a) the nonlinearity parameter $K$, b) the initial condition of $x$ and c) the slope $y_x$ of $ds(t)$.
\begin{equation}\label{lcnxy}
LCN=LCN(K,x,y_x) 
\end{equation}
The Lyapunov time is the inverse of the Lyapunov characteristic number:
\begin{equation}\label{tl}
t_L=1/LCN \footnote{For $K>K_c$ one can use the formula: \\ $t_L(K)=\frac{1}{LCN(K)-LCN(K_c)}$. The value of $LCN(K_c)$ is $\approx 0.1$ and for  $K>1$ the formula (\ref{tl}) is sufficiently accurate. }
\end{equation}

For values of $K$ below the critical value $K=K_c\approx 0.97$ the chaotic zones of the phase space do not communicate with each other. Therefore the calclulation of the $LCN$ is made locally and is different for each different chaotic region. For a given chaotic region the $ftLCN$ converges to the same value of $LCN$ independently of the initial conditions inside this region.
 
 \begin{figure*}
\centering
\includegraphics[scale=0.3]{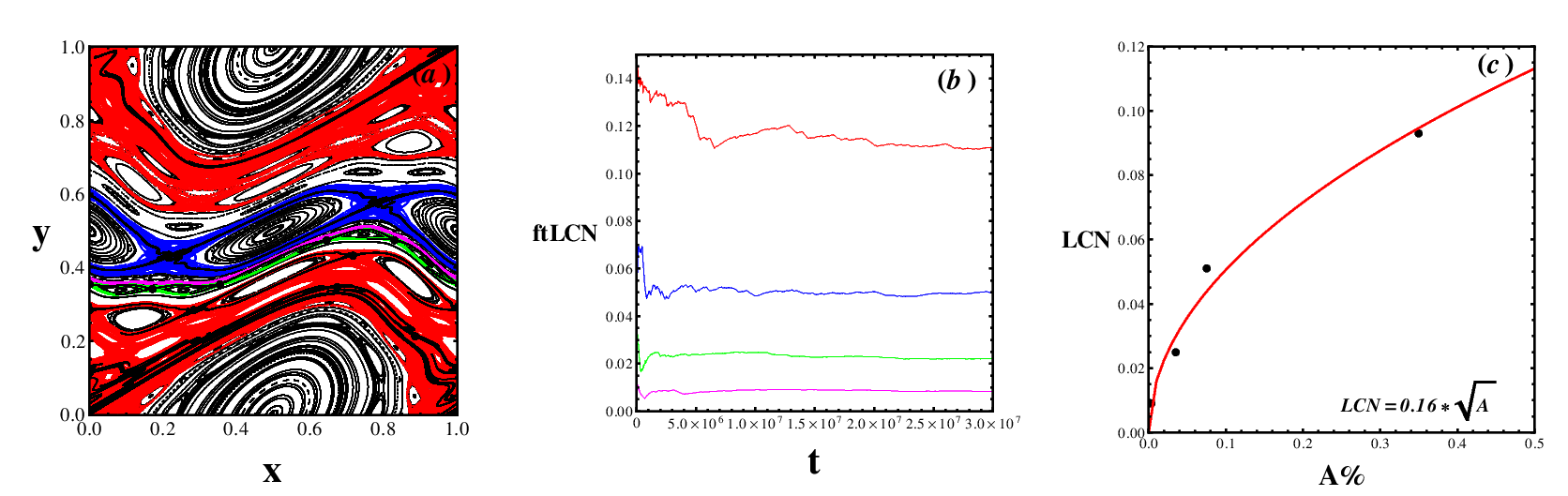}
\caption{ (a) The phase space of the standard map for the case $K=0.95$. The different chaotic regions (red,blue, green and magenta) do not communicate with each other. Black curves give the unstable manifolds of the simplest unstable periodic orbits in each region. (b) The evolution of the $ftLCN$ for each chaotic region derived from eq. (\ref{lcn}) (the colors of the curves are related to the colors of the chaotic areas). (c) $LCN$ as a function of the percentage of the available chaotic area $A$ is fitted approximately by the equation $LCN=c\sqrt{A}$, with $c=0.16$ for $K=0.95$.} \label{k095}
\end{figure*}

In Fig. \ref{k095}a we plot the phase space for $K=0.95$. The different chaotic regions (presented with different colors) do not communicate with each other and have different $LCNs$ as we see in Fig.  \ref{k095}b (the colors of the curves are related to the colors of the chaotic areas in Fig. \ref{k095}a). The values of the $LCNs$ are correlated with the sizes of the corresponding chaotic area. The $ftLCN$ needs a very long time in order to converge to each final value of $LCN$ when $K$ is near the critical value $K_{cr}\approx 0.97$. The black curves in Fig. \ref{k095}a correspond to  the unstable manifolds of the simplest periodic orbits of each region. A possible explanation of the correlation of the $LCN$ with the chaotic area is the fact that the local stretching number $\alpha$ depends on the local slope of the deviation  vector $y_x=dy/dx$ (see eq. (\ref{alfan})). The phenomenon of stickiness of the chaotic orbits along the unstable manifolds of each chaotic area (see \cite{2010CeMDA.107...77C}, \cite{2010IJBC...20.2005C}) connects the local $y_x$ of each chaotic orbit with the angles of the eigenvectors of the unstable periodic orbits of each chaotic subregion.  The angles of the eigenvectors of the unstable periodic orbits of the larger chaotic area (red) have greater values than the ones of the intermediate chaotic area (blue) and the angles of the eigencvectors of the blue area have greater values than the ones of the green area etc. In general, the mean slope of the unstable manifolds for each periodic orbit with the same multiplicity in all these areas is correlated with the size of the corresponding area. 
\begin{figure}
\centering
\includegraphics[scale=0.3]{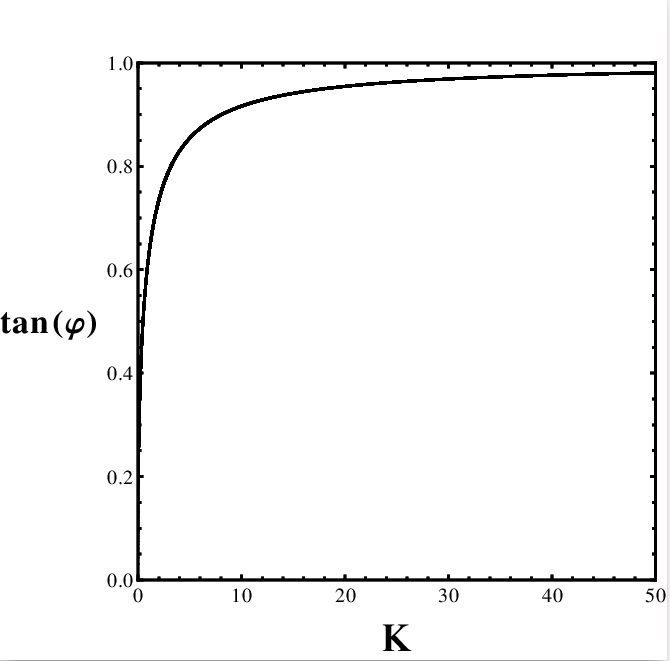}
\caption{The tangent of the angle $\phi$ of the eigevector $tan(\phi)$ of the unstable periodic orbit of simple multiplicity at the origin ($x=0$, $y=0$) as a function of $K$. } \label{eig}
\end{figure}
In Fig. \ref{k095}c the value of $LCN$ is plotted as a function of the percentage of the covered chaotic area $A$ (black dots). The red curve corresponds to the analytic formula:
\begin{equation}\label{lcnanal}
LCN=c \sqrt{A}
\end{equation}
with $c=0.16$, for $K=0.95$,
 which seems to describe well the numerical points. We have verified that this is a general formula presenting a correlation between the $LCN$ and the available chaotic area, with different values of $c$ for every $K$, when $K<K_c$, i.e. when the chaotic regions of the phase space do not communicate with each other.
  
\begin{figure*}
\centering
\includegraphics[scale=0.3]{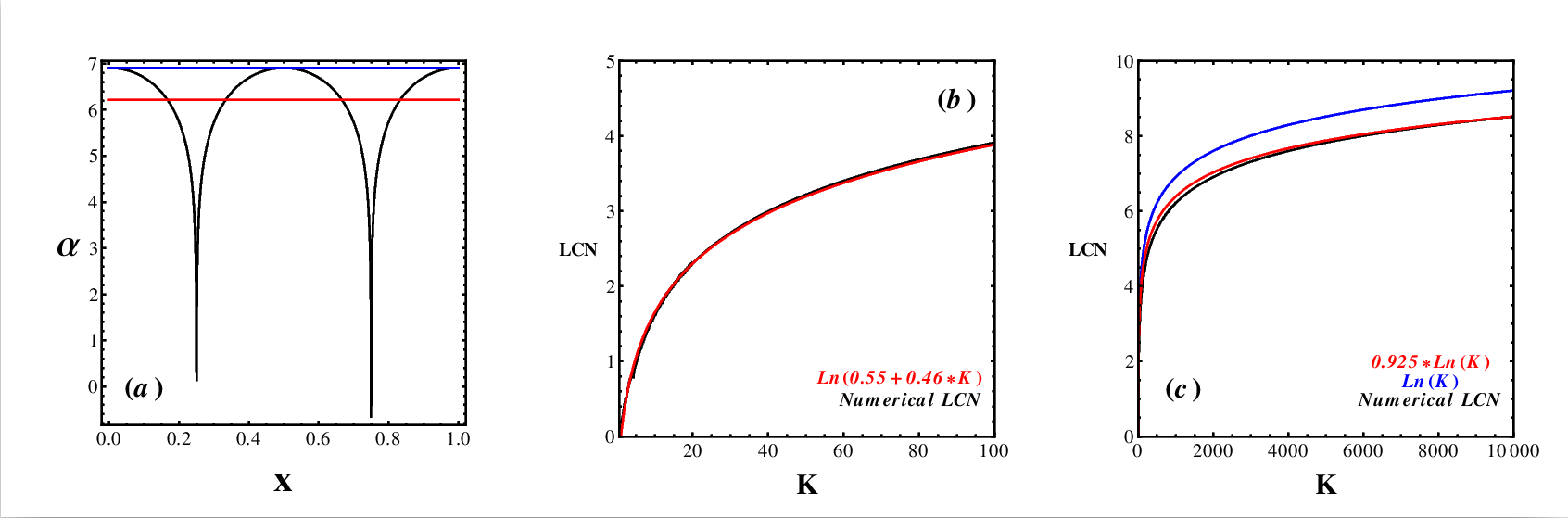}
\caption{ (a) The stretching number $\alpha$ as a function of $x$ for $K=1000$ (black curve) derived from eq. (\ref{alfan}) setting $y_x=1$. The red l52ine corresponds to the mean value $\langle \alpha\rangle$ and the blue line to the maximum value of $\alpha$ for $\cos(2 \pi x)=1$ and $y_x=1$ (which is valid for $K$  $\rightarrow \infty$). (b) The numerical value of $LCN$ as a function of $K$ calculated from eq. (\ref{lcnxy}) for the chaotic region of the phase space (black curve). The  function $LCN=\ln(0.55+0.46K)$ (red curve) is a good fit for relatively small values of $K$. (c) The numerical value of $LCN$ as a function of $K$ calculated from eq. (\ref{lcn}) for the chaotic region of the phase space (black curve). The blue curve corresponds to the simple function $LCN=\ln(K)$. The function $LCN=0.925 \ln(K)$ (red curve) coincides well with the numerical curve for large $K$. }\label{streth}
\end{figure*} 

When $K>K_c$ there is a unique chaotic sea that covers most of the phase space.  For these cases we want to derive an analytical relation of the $LCN$ as a function of the nonlinearity parameter $K$. For large values of $K$ the slope $y_x$ in eq. (\ref{alfan})  becomes equal to 1 as the tangent of the angle of the eigevector of the simple unstable periodic orbit ($x=0$, $y=0$) of multiplicity 1, converges quickly to 1 as a function of $K$ (see Fig. \ref{eig}). This means that the eigendirection of the unstable manifold of the periodic orbit of the origin ($x=0$ , $y=0$) becomes parallel to the diagonal of the phase space and its angle converges towards an angle of $45^o$ as $K\rightarrow \infty$. The same is true for all the manifolds of the unstable periodic orbits in the chaotic sea as they are forced to move parallel to each other. (The slope $y_x$ of the deviation vector of the chaotic orbits becomes equal to the slope of the unstable manifolds, locally, due to the phenomenon of stickiness along the manifolds \cite{2010CeMDA.107...77C}).

In Fig. \ref{streth}a we plot the stretching number $\alpha$ as a function of $x$, for $K=1000$, using equation (\ref{alfan}) and setting $y_x=1$ (black curve). The red  line corresponds to the mean value $\langle \alpha\rangle$ while the blue line gives the maximum value of $\alpha$ when $\cos(2 \pi x)=1$. The minima of the function $\alpha(x)$ correspond to initial conditions of stable or unstable periodic orbits with the minimum H$\acute{e}$non stability index (see \cite{2002ocda.book.....C}) for the specific $K$.

 In Fig. \ref{streth}b
 we plot the numerical value of the $LCN$ calclulated from eq. (\ref{lcn}) for a sufficient large number of iterations in order to secure the convergence of the $ftLCN$ to the value of $LCN$, as a function of $K$. The red curve is a fitting curve which corresponds to the analytical formula $LCN = \ln({0.55+0.46 K})$ and fits well the numerical curve for relatively small values of $K$ (but $K>K_c$).
For large values of $K$ the value of $y_x$ in eq. (\ref{alfan}) converges to 1 and the maximum value of the stretching number $\alpha$, (for $\cos(2 \pi x)=1$) is given by the formula $\alpha_{max}=\ln{K}$.

 In Fig. \ref{streth}c
 we plot the numerical value of the $LCN$ calclulated from eq. (\ref{lcn}) (black curve). The blue curve corresponds to the maximum value of the stretching number $\alpha$ which gives the function
$ LCN=\ln{K}$. 
Finally  we multiply $\ln{K}$ by a  coefficient 0.925 which gives the corrected function $LCN(K)$ (red curve in Fig. \ref{streth}c) and coincides well with the numerical curve for large values of $K$ (black curve). Therefore the analytical relations of the $LCN$ as a function of $K$ (for $K>K_c$) are: 

\begin{eqnarray}
\nonumber LCN=\ln(0.55+0.46 K), ~~~for~ small~ K\\
 LCN=0.925\ln{K}, ~~~for~ large~ K
\end{eqnarray}

\begin{figure}[hbt]
\centering
\includegraphics[scale=0.3]{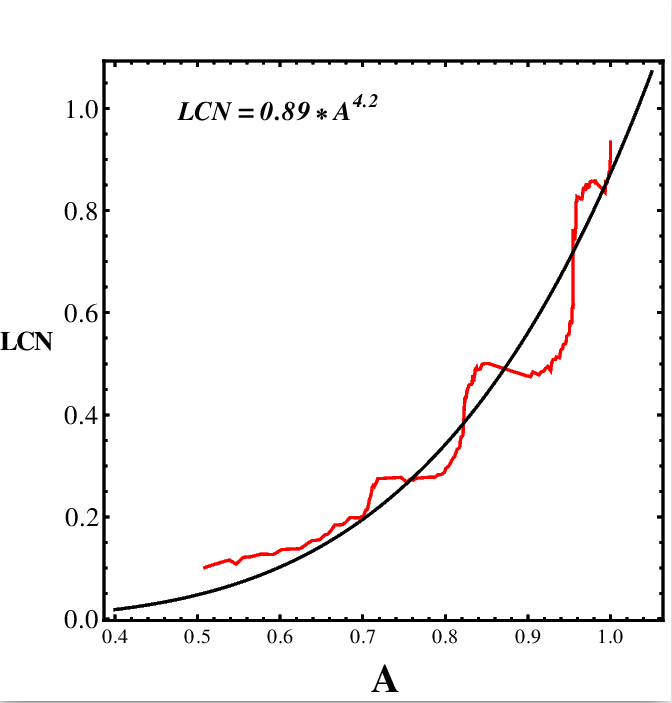}
\caption{$LCN$ as a function of the percentage of the available chaotic area $A$ of the phase space for $1<K<7$ (red curve). Each point corresponds to a specific value of $K$. The black curve is the best fit given by the relation: $LCN=0.89 A^{4.2}$.} \label{lcnak}
\end{figure}

Now we want to determine the relation between $LCN$ and the percentage of the availbale chaotic area $A$, for $K>K_c$. 
For this purpose we calculate the $LCN$ and the available chaotic area $A$ for values of the perturbation parameter: $1<K<7$ (after the value $K=7$ the percentage of the chaotic area is very close to $100\%$). 
In Fig. \ref{lcnak} we plot the $LCN$ as a function of the available chaotic area $A$ (red curve). Each point corresponds to the same value of $K$. 
A power law fit of this curve (black) has the following form: 
\begin{equation}\label{lcnakc}
LCN=cA^{4.2}
\end{equation}
where $c=0.89$. 
The deviations of the numerical (red) curve from the analytical (black) curve are due to the values of the area $A$ which have some abrupt changes as $K$ increases. This is due to the breaking of the last KAM curves surrounding islands of stability, increasing abruptly the available chaotic area in specific resonances (see \cite{2005IJBC...15.2865C}).  

\begin{figure*}
\includegraphics[scale=0.2]{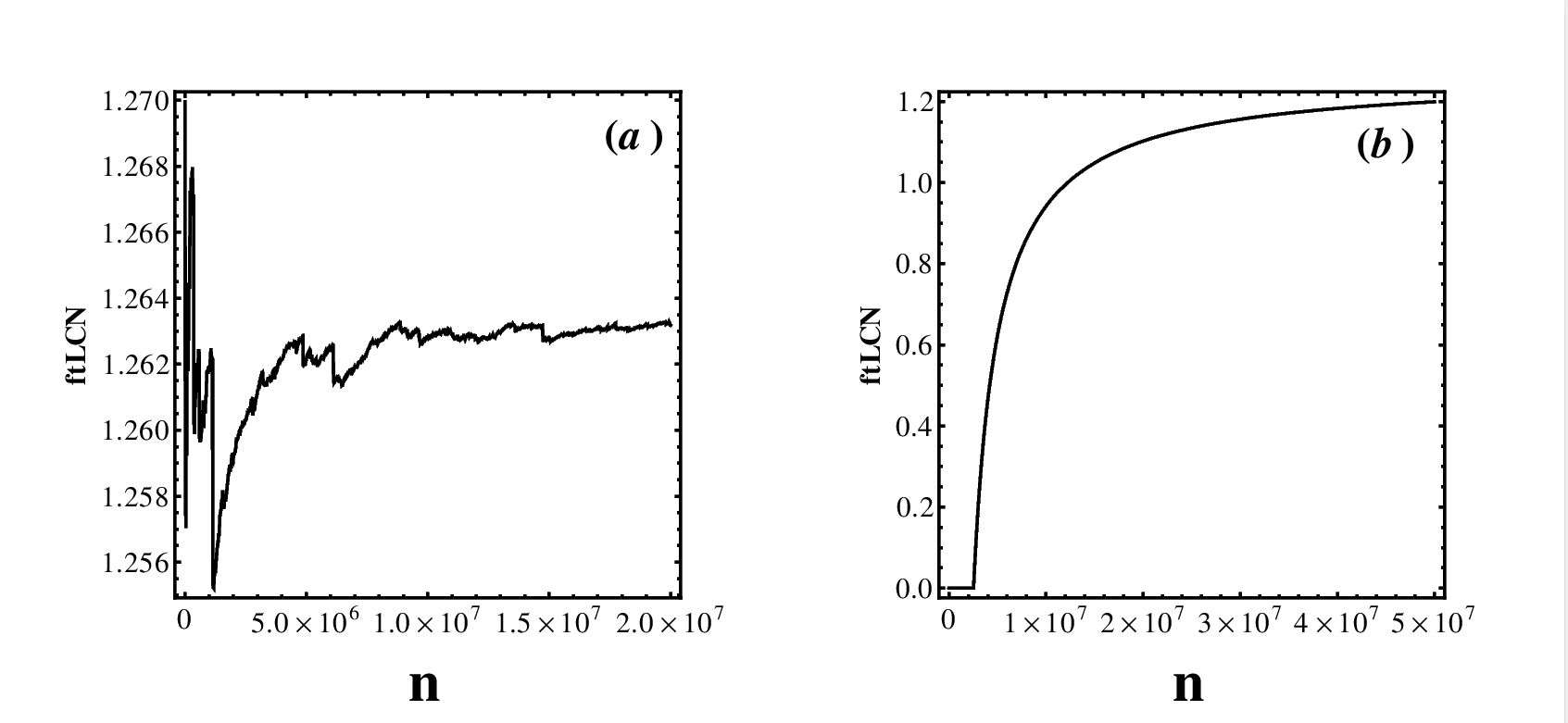}
\caption{(a) Typical dependence of the $ftLCN$ as a function of the number of iterations for an initial condition inside the large chaotic sea. (b) The $ftLCN$ as a function of the number of iterations for an initial condition inside an extreme sticky region (blue dot in Fig. \ref{zoom}e). } \label{ftlcn}
\end{figure*}

We point out here that the time of convergence of the $ftLCN$ to its final value of $LCN$ depends strongly on the initial conditions and can be very long in cases of extremely sticky regions around islands of stability. An example is given in Fig. \ref{ftlcn} where the evolution of the $ftLCN$ is calculated for two different initial conditions inside the chaotic area of the phase space for $K=6.908745$. (This case is studied in detail in section \rom{6}).  
In Fig. \ref{ftlcn}a we plot the time evolution of the $ftLCN$ for an initial condition inside the large chaotic sea and far away from the islands of stability. The $ftLCN$ has well converged after $10^7$ iterations. On the other hand, in Fig. \ref{ftlcn}b 
where the initial condition is inside the extreme sticky zone around an accelerator mode island (blue dot in Fig. \ref{zoom}e) the $ftLCN$ has not well converged even after $5 \times 10^7$ iterations. In this latter case, there exists an initial transient  period where the $ftLCN$ is close to zero corresponding to the stickiness around the accelerator mode island. Then we observe a very slow convergence towards the corresponding $LCN$ of the chaotic region. 

From Fig. \ref{streth}c we conclude that the Lyapunov time $t_L$ (which is the inverse of the $LCN$) is very small (smaller than 1),  and is a decreasing function of the nonlinearity parameter $K$.

\section{The Poincare recurrence time}
\begin{figure*}[hbt]
\centering
\includegraphics[scale=0.3]{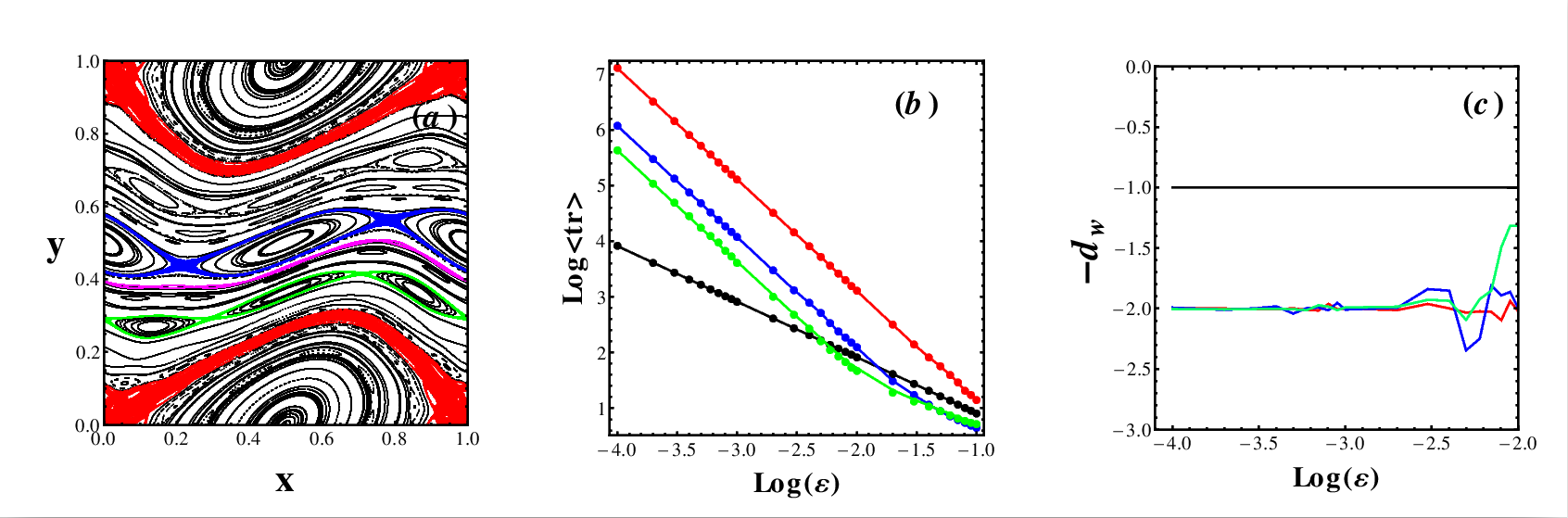}
\caption{(a) The phase space of the standard map for $K=0.8$. The different colors correspond to different chaotic domains that do not communicate with each other. (b) The logarithm of the Poincar$\acute{e}$ Recurrence time $\langle t_r\rangle$ as a function of the logarithm of the size of the box $\epsilon$ for the different chaotic areas of (a) (the curves have the same color with their corresponding chaotic domains). The black curve corresponds to initial conditions inside an island of stability. (c) The fractal dimensions $d_w$ (equal to the absolute value of the slope of the curves in (b)) as  functions of the logarithm of the size of the box $\epsilon$. }\label{dim} \end{figure*}

As we have already mentioned in the introduction the standard map has a mixed phase space both with chaotic and regular motions. Measuring the Poincar$\acute{e}$ recurrence times can give us information about the 
fractal (Hausdorff or box) dimension of the invariant subset (see \cite{afraimovich2006fractal} for a review).

The dynamics near the boundary of an island of stability is complicated due to  phenomena of
stickiness. This circumstance influences almost all important probability distributions such as the distribution of distances, exit times, recurrences, moments, etc. The main feature of all such
distributions is that they do not correspond to either Gaussian or Poissonian (or
similar) processes. This  manifests itself by the presence
of powerlike tails in the asymptotic limit of distributions for long-time and small-space scales (see the discussion below).

We study first the Poincar$\acute{e}$ recurrence times for a value of $K$ below the critical value $K_c$. When the different chaotic regions do not communicate with each other the recurrence times are correrated with the separated available chaotic area.   

In Fig.\ref{dim}a we plot the phase space for $K=0.8$ and we mark with different colors four different chaotic domains that do not communicate with each other. We calculate the mean Poincar$\acute{e}$ recurrence time for each chaotic domain for a number of $10^5$ initial conditions inside a square of size $\epsilon$ as a function of $\epsilon$ (Fig. \ref{dim}b) for the different chaotic domains of Fig.\ref{dim}a, in a logarithmic scale. For $\epsilon\rightarrow 0$, the recurrence times become astronomically large which reflects the original thought by Poincar$\acute{e}$ \cite{1893poincare}.  The curves have the same color with their corresponding chaotic domains. The black curve corresponds to regular motion with initial conditions inside an island of stability. All the curves corresponding to the chaotic regions have the same slope but the values of the mean recurrence times are correlated with the area of these domains, because for bigger chaotic domains the orbit has to cover statistically a larger area before returning in the same initial square. 

In Fig.\ref{dim}c we plot the absolute values of the slopes of the curves of Fig.\ref{dim}b as  functions of the logarithm of the size of the initial square $\epsilon$. These values converge to the fractal dimension $d_w$, for $\epsilon \rightarrow 0$ (see \cite{afraimovich2006fractal}). Actually, the slope for the case of chaotic regions give a fractal dimension $d_w=2$ which is identical with the dimension of the mapping, while the regular orbit gives a fractal dimension $d_w=1$, as the KAM curves inside the islands of stability are objects of dimension 1.

Buric et al. \cite{2003JPhA...36L.209B}  find that the spectrum of Poincar$\acute{e}$ recurrence times $F(t)$ (where $F(t)=N(t_{rec} \pm dt_{rec}) / N_{total} $, with $N(t_{rec} \pm dt_{rec})$ the number of initial conditions in a specific box of dimension $\epsilon$ having Poincar$\acute{e}$ recurrence times inside the interval $t_{rec} \pm dt_{rec}$ and  $N_{total}$ is the total number of initial conditions inside the box) for the standard map exhibits
two distinct limits: a weak-coupling limit with an inverse power
law decay and a chaotic strong-coupling limit with an exponential decay. An inverse power law was given first by Chirikov and Shepelyansky \cite{1984PhyD...13..395C}, while the distinction between exponential and power-like distributions was first made by Zaslavsky et al. \cite{zaslavsky1997self}. Long tails in distributions were first introduced in the context of modern statistical physics, by Montroll and Shlesinger \cite{1983JSP....32..209M}.  On the other hand, the deviation from the exponential decay in the case of weak chaos (or stickiness phenomena) is emphasized in a number of papers (e.g. \cite{zaslavsky2002chaos}, \cite{2004Chaos..14..975A}).   Buric et al. \cite{2003JPhA...36L.209B} give a relation for the spectrum $F(t)$ for domains
where chaotic regions coexist with integrable structures, with a superposition of an exponential and a power law decay of the form:
\begin{equation} \label{recft}
F(t)\simeq (1-p) \exp \Big(-t \frac{1-p}{1-P}\Big) +\frac{p}{2} \Big(\frac{P}{pt}\Big)^2
\end{equation}
where $p$, $P$ are  suitable constants fitting each curve. 
 We give here a slightly more general relation where the power of the inverse power law can be different from 2:
\begin{equation} \label{recftn}
F(t)\simeq (1-p) \exp \Big(-t \frac{1-p}{1-P}\Big) +\frac{p}{2} \Big(\frac{P}{pt}\Big)^a
\end{equation}

Such a law is found numerically to occur in area-preserving maps at the boundary of the
mixing and integrable components, i.e. in areas of stickiness around islands of stability, but also in chaotic regions that are limited by closed KAM curves and do not communicate with each other. Such cases of isolated chaotic regions can be found for values of $K$ below the critical value $K_c \approx0.97$.  

\begin{figure*}
\centering
\includegraphics[scale=0.4]{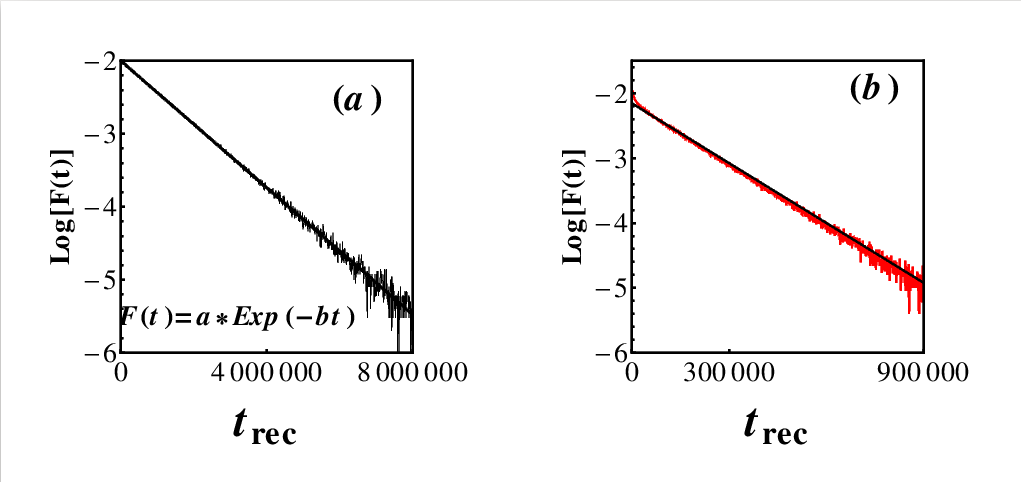}
\includegraphics[scale=0.4]{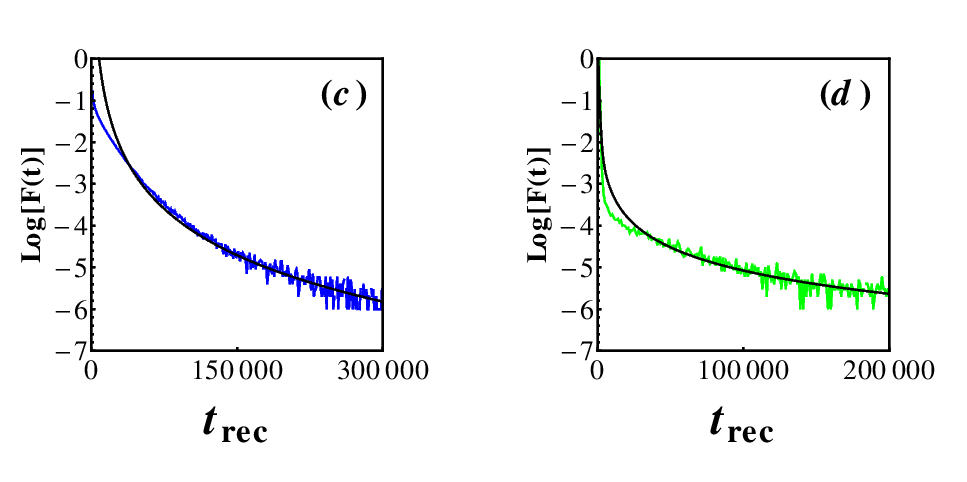}
\caption{(a) The distribution of the Poincar$\acute{e}$ recurrence times (spectrum) $F(t)$ for the main chaotic domain for $K=6.0$ presents an exponential decay  (b) the spectrum $F(t)$ for the red chaotic region of Fig. \ref{dim}a ($K=0.8$) is close to exponential  (c) the spectrum $F(t)$ for the blue chaotic region of Fig. \ref{dim}a deviates from the exponential decay and can be approximately fitted by the formula (\ref{recftn}) (d) the spectrum $F(t)$ for the green chaotic region of Fig. \ref{dim}a fitted by the formula (\ref{recftn}).  }\label{rec}
\end{figure*}

In Fig. \ref{rec} we plot the distribution of the Poincar$\acute{e}$ recurrence times $F(t)$ of an ensemble of $10^6$ initial conditions inside a small area of dimension $\epsilon = 10^{-3}$ in different domains of the phase space. Fig. \ref{rec}a corresponds to a case of strong chaos, i.e. the initial conditions are taken insidewhere the main chaotic domain for $K=6.0$. Here the orbits can cover the whole available chaotic part of the phase space. The spectrum is given in a semi-logarithmic plot and presents an  exponential behavior. The same is true in Fig. \ref{rec}b  for the case of the red chaotic domain of Fig. \ref{dim}a where  $K=0.8$ and the exponential fitting is suitable apart from a small part of the curve corresponding to small recurrence times. Moreover, the distribution $F(t)$ drops off at a small value of $t_r$ (it does not appear in the figure). In fact the recurrence time $t_r$ cannot be smaller than a minimum value (see \cite{1996CeMDA..63..189C}). The minimum recurrence time has been found in  particular cases \cite{2003CeMDA..85..105P}, \cite{2004CeMDA..88..163C}. Near the minimum recurrence time ($t_{r_{min}}$) the formula (\ref{recftn}) is not valid and the curve $F(t)$ goes to zero as $t_{r}$ goes below $t_{r_{min}}$ in Fig. \ref{rec}.

On the other hand for the blue and the green chaotic regions of Fig. \ref{dim}a, the spectrum of the Poincar$\acute{e}$ recurrence times deviates from the exponential fitting and presents long power-law tails (Figs.~\ref{rec}c,d). The fitting in this case can be best approximated by the relation (\ref{recftn}) with $a=3.7$ in Fig. \ref{rec}c and $a=1.8$ in Fig. \ref{rec}d. In general, the smaller the area of the chaotic region is, the greater is the deviation from the exponential decay.  The same relation describes the distribution of  Poincar$\acute{e}$ recurrence times for $K>K_c$, for initial conditions in regions very close to islands of stability due to phenomena of stickiness.  We give an example in section \rom{6} (Figs. \ref{distri}b,d).

\begin{figure}
\centering
\includegraphics[scale=0.25]{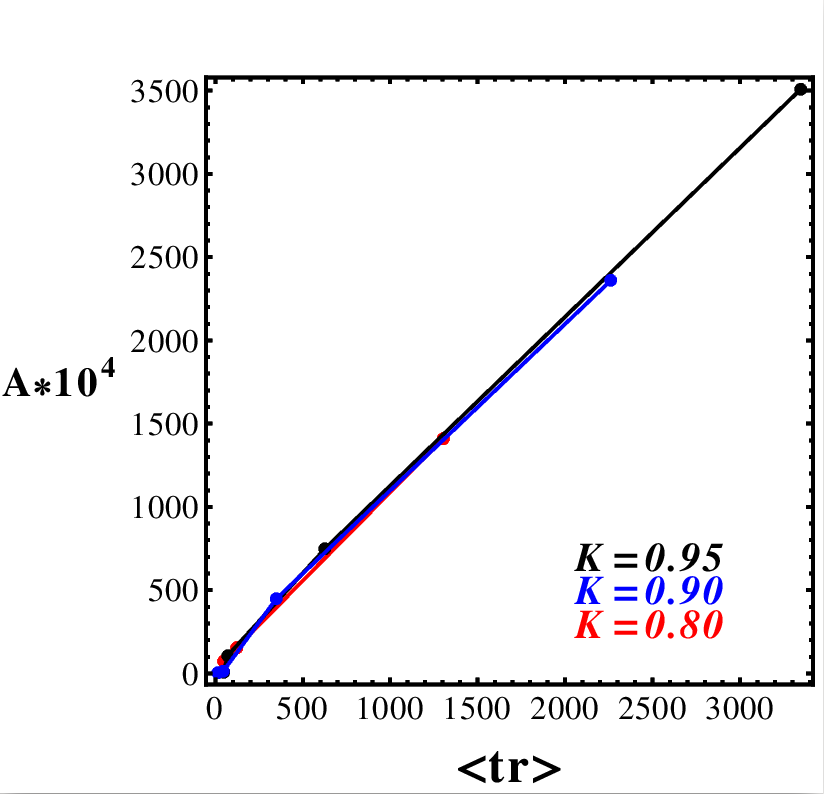}
\caption{The percentage of the available chaotic area $A$ (multiplied by a factor $10^4$) for three discrete chaotic regions as a function of the mean Poincar$\acute{e}$ recurrence time $\langle t_r\rangle$, for a box size $\epsilon=10^{-2}$ and for  $K=0.8$ (red curve), $K=0.9$ (blue curve) and $K=0.95$ (black curve). The curves are close to the diagonal.} \label{atrkc}
\end{figure}

\begin{figure}
\centering
\includegraphics[scale=0.25]{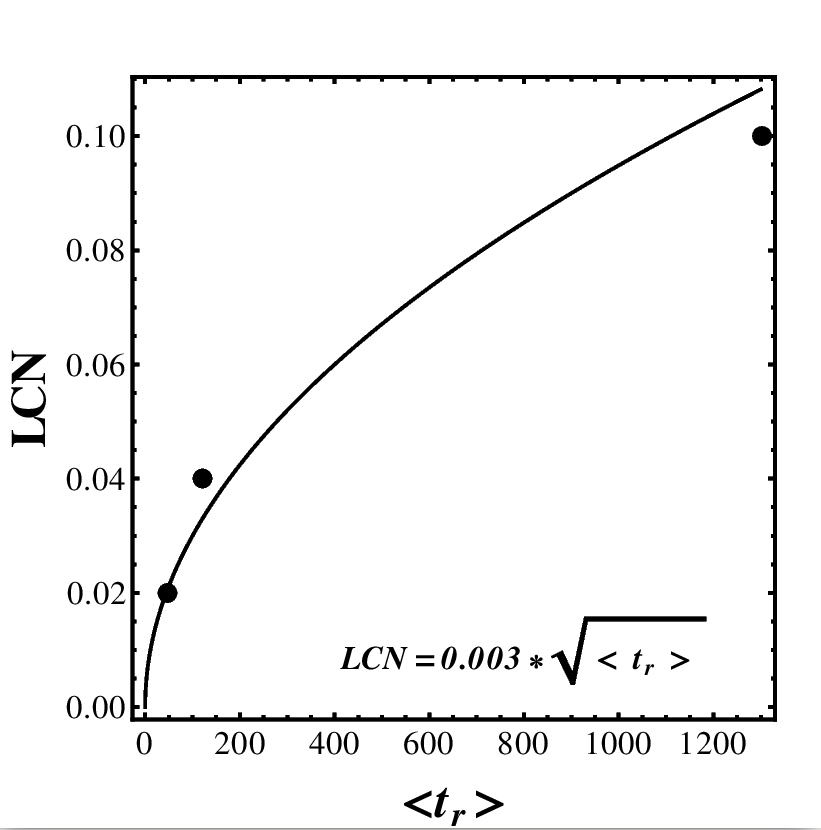}
\caption{The mean Poincar$\acute{e}$ recurrence time $\langle t_r\rangle$ is correlated with the LCN of the different chaotic regions, for $K=0.8$.} \label{atr1}
\end{figure}

In the previous section we have shown that the LCN is well correlated with the available chaotic area covered by a chaotic orbit. In this section we show that the mean Poincar$\acute{e}$ recurrence times $\langle t_r\rangle$ are correlated as well with the available chaotic area. 
When $K<K_c$ where the different chaotic regions do not communicate with each other, the mean Poincar$\acute{e}$ recurrence time $\langle t_r\rangle$ is related to the percentage of the chaotic area and the size of the box $\epsilon$ of the initial conditions by the following relation:
\begin{equation}\label{trakc}
\langle t_r\rangle=bA \epsilon ^{-2}
\end{equation}
where the coefficient $b$ is close to 1. 
In Fig. \ref{atrkc} we plot the percentage of the available chaotic area $A$ multiplied by a factor of $10^4$ for three discrete chaotic regions as a function of the mean Poincar$\acute{e}$ recurrence time $\langle t_r\rangle$, for a box size $\epsilon=10^{-2}$ and for $K=0.8$ (red curve), $K=0.9$ (blue curve) and $K=0.95$ (black curve).  It is obvious that the curves satisfy eq. (\ref{trakc}). 

Using eqs. (\ref{lcnanal}) and (\ref{trakc}) we derive a relation between the $LCN$ and the mean Poincar$\acute{e}$ recurrence time $\langle t_r\rangle$:
\begin{equation}\label{trlcn}
LCN=g\epsilon \sqrt{\langle t_r\rangle}
\end{equation}
where $g=g(K)$ is a coefficient depending on the nonlinearity parameter $K$. An example for three different chaotic regions in the case $K=0.8$, is shown in Fig. \ref{atr1} where $g \epsilon=0.003$.

\begin{figure*}[hbt]
\centering
\includegraphics[scale=0.25]{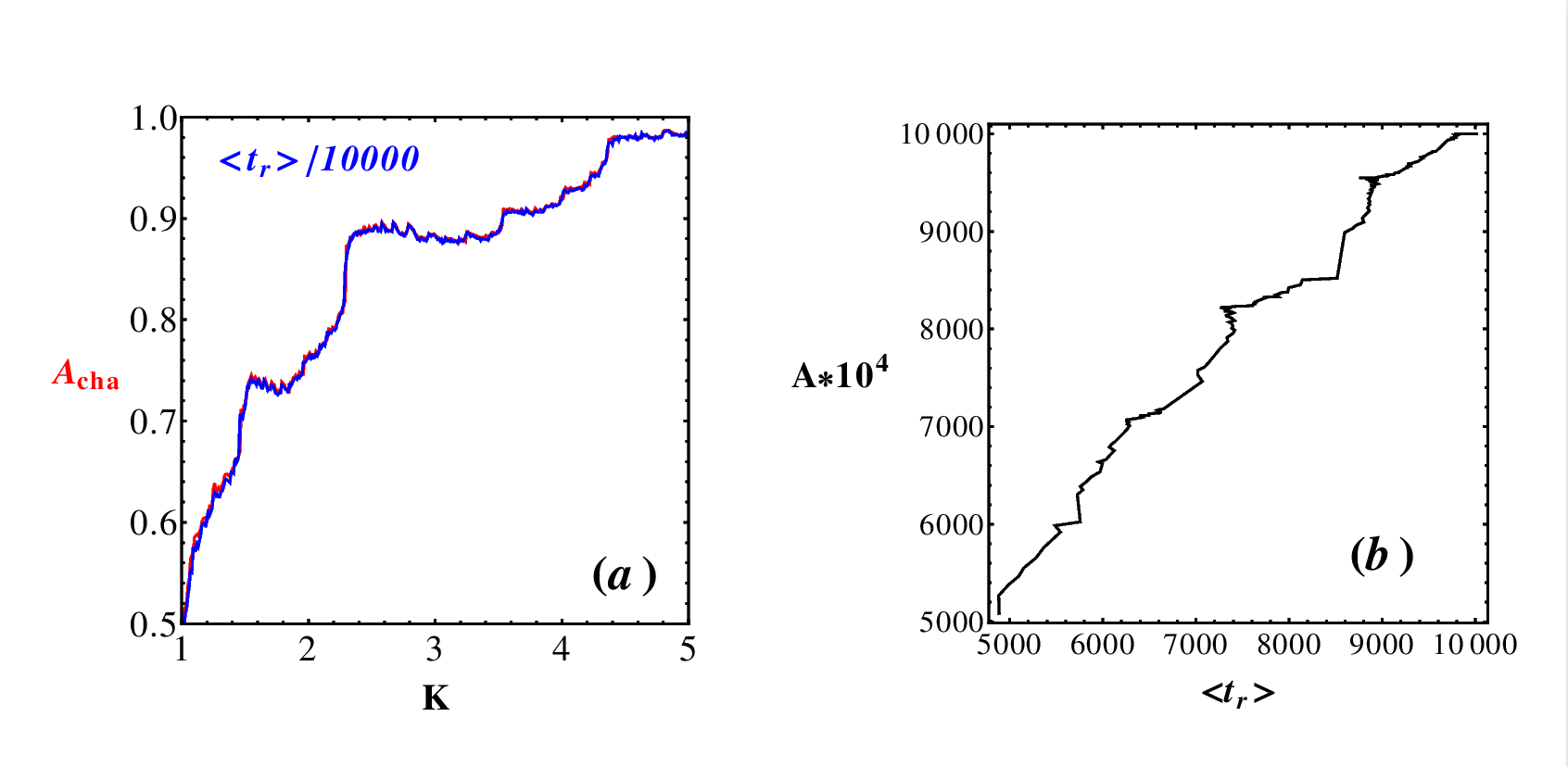}
\caption{(a) The red curve corresponds to the percentage of the chaotic area of the phase space $A$ as a function of the parameter $K$. The blue curve corresponds to the mean Poincar$\acute{e}$ recurrence time $\langle t_r\rangle$ inside the chaotic area in a box of dimension $\epsilon=10^{-2}$ divided by $10^4$. The correlation  between the two curves is extremely good. (b)  The percentage of the chaotic area  $A$ multiplied by a factor of $10^4$ as a function of the mean Poincar$\acute{e}$ recurrence time $\langle t_r\rangle$ is very close to the diagonal. } \label{atr}
\end{figure*}

\begin{figure}[hbt]
\includegraphics[scale=0.27]{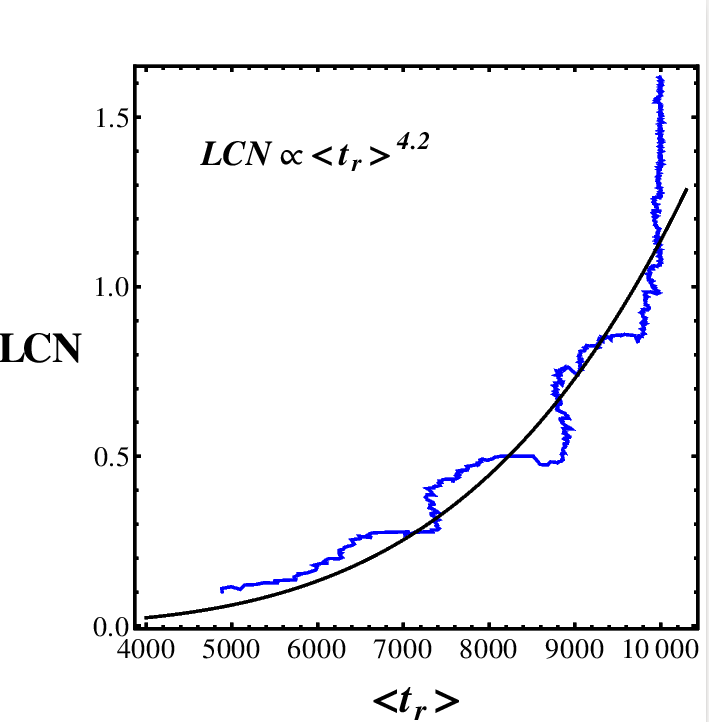}
\caption{The function $LCN=f(\langle t_r\rangle)$ for values of nonlinearity parameter in the range $K_c<K<7$ (blue curve). The power law fitting $LCN \propto \langle t_r\rangle^{4.2}$ (black curve) is a good approximation. For larger values of $K$ this power law is no longer valid because $\langle t_r\rangle$ tends to an almost constant value.  } \label{lcntrb}
\end{figure}

\begin{figure*}[hbt]
\includegraphics[scale=0.3]{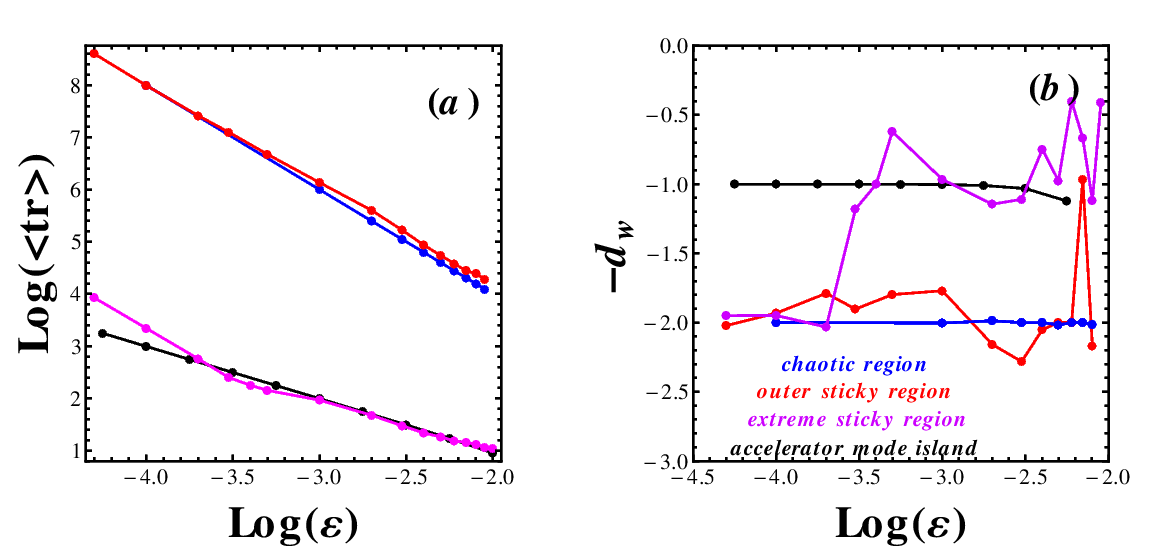}
\caption{(a) The logarithm of the Poincar$\acute{e}$ recurrence time $\langle t_r\rangle$ as a function of the logarithm of the size of the box $\epsilon$ for $K=6.608$. Black curve corresponds to the accelerator mode island, magenta to the extreme sticky region outside the accelerator mode island, red to the outer sticky region and blue to the chaotic region far away from the islands of stability  (b) The fractal dimensions $d_w$ (equal to the absolute values of the slope of the curves in (a)) as  functions of the logarithm of the size of the box $\epsilon$. The fractal dimension $d_w$ is related to the diffusion exponent $\mu$ through the relation $d_w= 2/ \mu $ (see text).} \label{poincst}
\end{figure*}

When $K>K_c$, although there is one united chaotic domain in the phase space of the standard map, in general there exist islands of stability that tend to be smaller and smaller as $K$ increases. Therefore, the percentage of the total available chaotic area $A$ is increasing with increasing $K$ and the mean Poincar$\acute{e}$ recurrence time $\langle t_r\rangle$ of the chaotic region is correlated with $A$. 

In Fig. \ref{atr}a we observe a striking correlation between $\langle t_r\rangle$ and $A$. The red curve corresponds to the percentage of the available chaotic area $A$ as a function of $K$, while the blue curve corresponds to the mean Poincar$\acute{e}$ recurrence time $\langle t_r\rangle$  of the chaotic domain as a function of $K$ for a box of size $\epsilon=10^{-2}$. The two curves present the same fractal shape and they are correlated even in extremely small scales. In order to compare the two quantities we have divided the values of the  mean Poincar$\acute{e}$ recurrence time $\langle t_r\rangle$ by a factor of $10^4$. Therefore, we verify that eq. (\ref{trakc}) is still valid for $K>K_c$, with $b=1$, i.e.
\begin{equation}\label{trakcn}
\langle t_r\rangle(K)=A(K) \epsilon ^{-2}
\end{equation}
 According to the eq. (\ref{trakcn}), the functional dependence of $\langle t_r\rangle$ on $K$ only enters through the percentage of the available chaotic area $A$, as the size $\epsilon$ of the selected box is independent of $K$. 

The percentage of the chaotic area $A$ tends asymptotically to 1 as $K$ increases, so for large enough values of $K$, the mean Poincar$\acute{e}$ recurrence time $\langle t_r\rangle$ is simply the inverse of the area of the selected box, i.e.: $\langle t_r\rangle=1/ \epsilon^2$, in a good approximation. This relation can be written in general:
\begin{equation}
\langle t_r\rangle=\frac{1}{\epsilon ^ {d_w}}
\end{equation}
where $d_w$ is the fractal dimension. For a 2-D mapping $d_w=2$  and is identical with the topological dimension. The same relation has been derived for some non-hyperbolic one dimensional mappings, (see  \cite{2011EPJB...82..219A}, \cite{2011karama}) where the authors have found the relation: $\langle t_r\rangle=1 / \epsilon$, as the topological dimension of the corresponding mappings is 1.

We have checked the validity of eq. (\ref{trakcn}) for a different size of the box, i.e. $\epsilon=10^{-3}$ and we have found that the two curves of Fig. \ref{atr}a coincide again if we divide $\langle t_r\rangle$ by a factor of $10^6$.
In Fig. \ref{atr}b we plot the percentage of the chaotic area  $A$ multiplied by a factor of $10^4$ as a function of the mean Poincar$\acute{e}$ recurrence time $\langle t_r\rangle$, for a box size $\epsilon=10^{-2}$. The curve is very close to the diagonal which means that the two quantities are equal in a very good approximation.

On the other hand, using eqs. (\ref{lcnakc}) and (\ref{trakc}) we derive the relation between $LCN$ and $\langle t_r\rangle$ for $K>K_c$:

\begin{equation}\label{lcntrkc}
LCN=q\langle t_r\rangle^{4.2}
\end{equation}
where the coefficient $q=q( \epsilon)$ depends on the size of the box $\epsilon$, selected for calculating the mean Poincar$\acute{e}$ recurrence time $\langle t_r\rangle$.  The function $LCN=f(\langle t_r\rangle)$ is shown in Fig. \ref{lcntrb}, where we see that the relation (\ref{lcntrkc}) is valid up to a perturbation parameter $K \approx 7$. However, for larger values of $K$ the percentage of the chaotic area $A$ goes very close to 1 and as a consequence the corresponding mean Poincar$\acute{e}$ recurrence time $\langle t_r\rangle$ converges as well to a constant value close to 10000 (which is $1/ \epsilon ^2$ in our case), without further significant increase with increasing $K$.

Extreme stickiness phenomena close to islands of stability can influence the mean Poincar$\acute{e}$ recurrence time. Such a case is presented in Fig. \ref{poincst}. The Poincar$\acute{e}$ recurrence time is calculated for $K=6.608$ for several initial conditions. In this case extreme stickiness phenomena occur near the border of the accelarator mode islands. 

In Fig. \ref{poincst}a we plot the logarithm of the Poincar$\acute{e}$ recurrence time $\langle t_r\rangle$ as a function of the logarithm of the size of the box $\epsilon$ for $K=6.608$. Black curve corresponds to initial conditions inside the accelerator mode island, magenta to the extreme sticky region close and outside the accelerator mode island, red to the outer sticky region and blue to the chaotic region far away from the islands of stability. 
In Fig. \ref{poincst}b we plot the fractal dimensions $d_w$, (equal to the absolute value of the slope of the curves of Fig. \ref{poincst}a) as  functions of the logarithm of the size of the box $\epsilon$.
The fractal dimension is related to the diffusion exponent $\mu$ by the relation (see \cite{2005drfd.book.....B}):

\begin{equation}
 d_w=2/ \mu  
\end{equation}
(see section \rom{5} for the definition of the diffusion exponent $\mu$). 

In Fig. \ref{poincst}b we observe that inside the accelarator mode islands the diffusion coefficient is $\mu=2$ (corresponding to anomalous diffusion and ballistic motion) and the corresponding fractal dimension is $d_w=1$. On the other hand in the chaotic region $\mu=1$ (corresponding to normal diffusion) and $d_w=2$. In the outer sticky region the diffusion coefficient has some variations around $\mu=1$, but for smaller $\epsilon$ it converges to $\mu=1$ (hence $d_w \approx 2$).

When the initial conditions are located inside the extreme sticky region of the accelerator mode islands the orbits present the same behavior with the KAM curves inside the islands of stability, i.e ballistic motion with diffusion exponent $\mu=2$ and fractal dimension $d_w=1$ for a rather big size of the box $\epsilon$, but for a smaller size of the box the diffusion converges into normal diffusion with $\mu=1$ and $d_w=2$. This happens because in the latter case the orbit has entered inside the large chaotic sea, where normal diffusion dominates (for more details see section \rom{5}).

\section{The Stickiness time}
The study of stickiness has expanded considerably in recent years. The first example of stickiness was provided by Contopoulos \cite{1971AJ.....76..147C}. The name ``stickiness'' was introduced by Shirts and Reinhardt \cite{1982JChPh..77.5204S} and after that many people worked on stickiness. Many related references are included in the paper of Contopoulos and Harsoula \cite{2010IJBC...20.2005C}. In that paper we discussed the various factors that affect the time scale of stickiness. 

The ``stickiness time'', also called ``escape time'', is the time required for the orbits to go far from their initial conditions into the large chaotic sea. After such a time the distribution of the points of these orbits becomes roughly uniform in the chaotic domain outside the islands of stability. 

However the stickiness time is different for different initial conditions. If an orbit starts closer to the boundary of an island (its last KAM curve) the stickiness time is longer and tends to infinity as the initial condition tends to reach this boundary (see \cite{2010IJBC...20.2005C}).

The stickiness times are much smaller than the recurrence times corresponding to the same region of initial conditions.
In a particular example with $K=6.608$ (Fig. 16 of \cite{2018PhRvE..97b2215H}) we considered 3 orbits starting close to an island of stability. The orbit $A$ has stickiness time $t_s(A)=5 \times 10^5$ iterations. The orbits $B$ and $C$ that are further away from the boundary of the island  have  stickiness times $t_s(B)=1.8 \times 10^5$
and $t_s(C)=1.2 \times 10^5$ respectively. The corresponding recurrence times are: $t_r(A)=4 \times 10^9$, $t_r(B)=5 \times 10^8$ and $t_r(C)=3.1 \times 10^7$ respectively. Therefore, the ratios of the recurrence times to the stickiness times are $t_r/t_s (A)=8 \times 10^3$, $t_r/t_s (B)=3.2 \times 10^3$ and $t_r/t_s (C)=2.6 \times 10^3$. The ratio $t_r/t_s $ is in general larger as we approach the boundary of the island.
A particular case of extremely long stickiness time is studied in the section \rom{6}, where we compare the distribution of the stickiness times and of the  Poincar$\acute{e}$ recurrence times.

\section{Stickiness and diffusion}

The stickiness phenomenon affects the diffusion of the orbits along the y-axis. The general diffusion is described by the formula: 

\begin{eqnarray} \label{diff3}
\langle(y-y_0)^2\rangle=D (K) n^\mu
\end{eqnarray}
where the $y$-component is calculated without the modulo 1 in eq. (\ref{stand}), $y_0$ is its initial value, $n$ is the number of iterations and $\langle\,\, \rangle$ denotes the mean value of an ensemble of initial conditions inside a small area of the phase space. Typical values of the effective diffusion $D( K )$ are given in \cite{2018PhRvE..97b2215H} and references within  and  $\mu$ is the diffusion exponent as it was introduced in \cite{zaslavsky2002chaos}. The diffusion exponent $\mu$ is defined in the long run limit, $n \rightarrow \infty$. 
For initial conditions in the chaotic domain, far away from the islands of stability, the diffusion is normal with $\mu=1$ (\cite{ishizaki1991anomalous}, \cite{zaslavsky1997self}, \cite{benkadda1997self}, \cite{manos2014survey}).  In this case, the
diffusion coefficient $D$ is a constant that depends only on $K$.
Inside the normal mode islands the diffusion exponent is equal to zero ($\mu=0$) and inside the accelerator mode islands the diffusion exponent is $\mu=2$. In the sticky zone around a normal island of stability the diffusion exponent is close to $\mu=0$ for some time and then it converges to $\mu=1$ (see Fig.4 of \cite{2018PhRvE..97b2215H}). On the other hand the diffusion exponent around an accelerator mode island stays close to $\mu=2$ for some time and after a crossover it finally converges to $\mu=1$ (Figs. 5b,c and 12b of \cite{2018PhRvE..97b2215H})). 

In the past, many people claimed that the diffusion exponent of the chaotic part of the phase space, in the case where accelerator mode islands exist, is between 1 and 2. In fact Venegeroles in \cite {2009PhRvL.102f4101V} quotes a number of papers for Hamiltonian systems with bounded phase space where the authors have calculated the diffusion exponent $\mu$ and concluded that the average value of $\mu$ outside the accelerator islands is close to $\mu=1.5$. However all these calculations were made for times of order of $10^6$. On the other hand we have shown in  \cite{2018PhRvE..97b2215H} that for longer times the diffusion exponent $\mu$, for initial conditions outside the accelerator mode islands, always tends to 1. In our examples we have found that orbits starting close to an accelerator mode island are dragged by these islands for times of the order of $10^6$, but later on they detach themselves from the accelerator islands and they enter the large chaotic sea tending to a diffusion exponent $\mu=1$.

 Similar results were found by Das and Gupte in \cite{2017PhRvE..96c2210D} who calculated the distributions of the diffusion exponents $\mu$ in a 3-dimensional mapping (extending their calulations up to $10^{10}$ iterations). They found only two main values of $\mu$ for the whole phase space, namely $\mu=1$ (normal diffusion) and $\mu=2$ (ballistic motion).

\begin{figure*}
\centering
\includegraphics[scale=0.23]{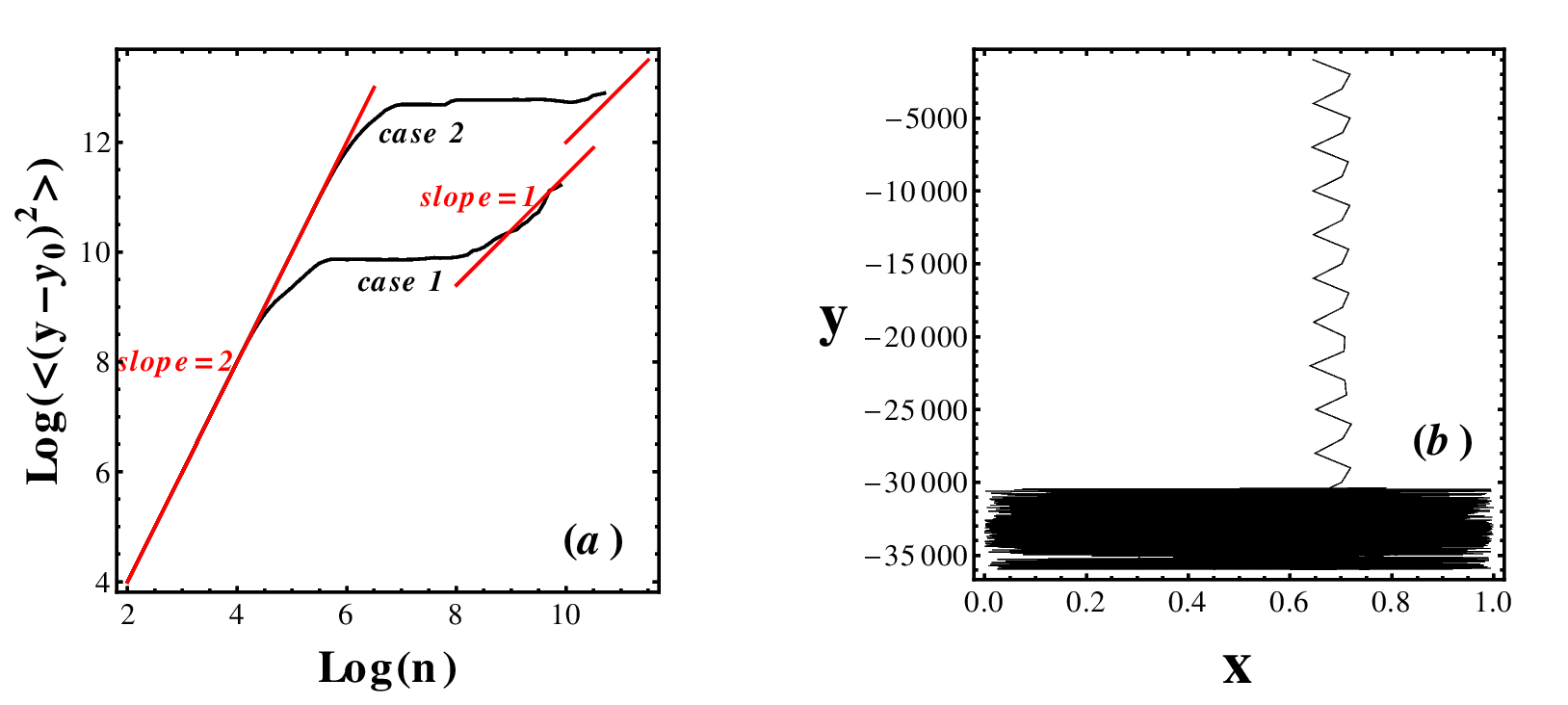}
\caption{ (a) The diffusion exponent $\mu$ (the slope of the time evolution of  $\langle y-y_0\rangle^2$ in logarithmic scale) calculated for two initial conditions: (1) in the vicinity of the sticky region (red dot in Fig. \ref{zoom}c), called ``case 1'' and (2) in the vicinity of the sticky region (blue dot in Fig. \ref{zoom}e), called ``case 2''. The diffusion is anomalous ($\mu=2$) during the stickiness time around the accelerator mode islands. Then in both cases the diffusion exponent $\mu$ drops and remains close to $\mu=0$ for a long time span and finally it converges to normal diffusion with $\mu=1$. (b) The evolution of the orbit corresponding to the ``case 1''. While the orbit executes ballistic motion the x-component stays localized and spreads in the whole phase space at the end of this motion.  }\label{dif}
\end{figure*}

In the present paper we give two examples of orbits in a case of extreme stickiness considered by Zaslavsky et al. \cite{zaslavsky1997self}, namely for $K=6.908745$ where a hierarchy of islands is presented in the phase space (see Fig. \ref{zoom} in the section \rom{7}). In this case we calculated the diffusion exponent $\mu$ for two different groups of orbits, one at the  level of Fig. \ref{zoom}c (initial conditions around the red dot) and the other at a deeper level of stickiness (initial conditions around the blue dot of Fig. \ref{zoom}e).
The first group of orbits corresponds to ``case 1'' of Fig. \ref{dif}a. It gives an average diffusion exponent $\mu=2$ for a little less than $10^5$ iterations, then it shows a crossover $\mu \approx 0$ up to $10^8$ iterations and afterwards the diffusion exponent converges to the value $\mu=1$ manifesting normal diffusion. The second group of orbits corresponds to ``case 2'' and gives  an average diffusion exponent $\mu=2$ for about $10^6$ iterations, then    
$\mu \approx 0$ up to $10^{10}$ iterations and finally $\mu$ tends to the value $\mu=1$. 
   
An explanation of this behavior of $\mu$ is provided in Fig. \ref{dif}b. Orbits close to the boundary of an accelerator mode island
are dragged close to this island by a ballistic motion, and $|y-y_0|$ increases almost linearly, thus $\mu=2$, for a time span equal to the stickiness time. While the orbits execute ballistic motion, the x-component stays localized in the phase space. At the end of the ballistic motion the orbits detach themselves from the boundary of the island and the x-component is spread in the whole phase space, while the y-component does not vary considerably for a rather long time span (which corresponds to $\mu=0$). Finally the orbits enter inside the large chaotic sea and the y-component has a mild increase corresponding to normal diffusion with $\mu=1$.

\section{Comparison of the characteristic times in an extreme stickiness case}

\begin{figure*}
\includegraphics[scale=0.4]{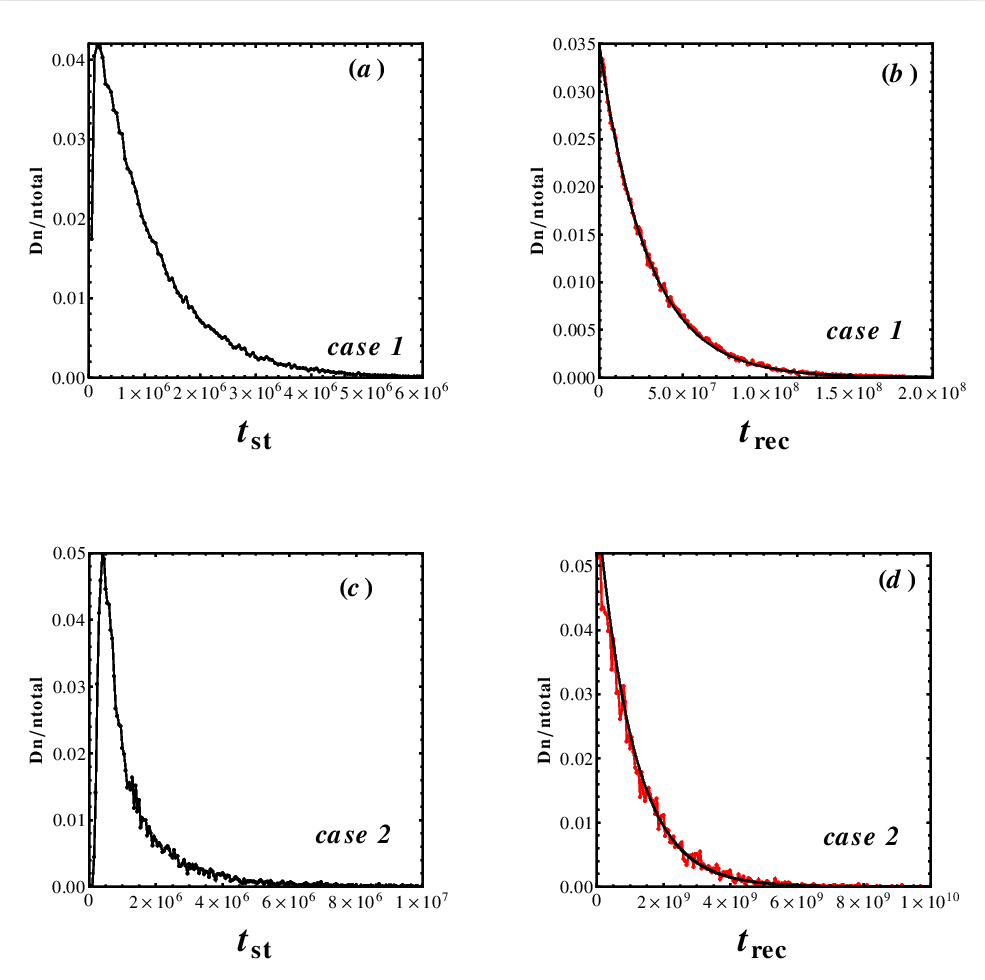}
\caption{(a) The distribution of the stickiness (or escape) times $t_{st}$ in the vicinity of the red dot of Fig. \ref{zoom}c (case 1). (b) The corresponding distribution of the Poincar$\acute{e}$ recurrence times $t_{rec}$. The black curve is a fitting using eq. (\ref{recft}). (c) The distribution of the stickiness (or escape) times $t_{st}$ in the vicinity of the blue dot of Fig. \ref{zoom}e (case 2). (b) The corresponding distribution of the Poincar$\acute{e}$ recurrence times $t_{rec}$. The black curve is a fitting using eq. (\ref{recft}).}. \label{distri}
\end{figure*}

As described in the previous section, sticky orbits with initial conditions close to the islands of stability stay there for times compared to the stickiness (or escape) time and then they enter inside the large chaotic sea. But after a Poincar$\acute{e}$ recurrence time they reenter inside the sticky region.  This reentry will happen for infinitely many times and one may claim that the diffusion exponent $\mu$ should converge to a value between 1 and 2. But this does not really happen for two reasons:
 
  (a) The Poincar$\acute{e}$ recurrence times are much longer than the stickiness times by orders of magnitude inside every region of stickiness. This is obvious in Fig. \ref{distri} where we plot the distributions of the stickiness (or escape) times and the Poincar$\acute{e}$ recurrence times for $K=6.908745$. Figs. \ref{distri}a,b correspond to the ``case 1''  (around the red dot in Fig. \ref{zoom}c) and Figs. \ref{distri}c,d correspond to the ``case 2''  (blue dot in Fig. \ref{zoom}e). We see that the Poincar$\acute{e}$ recurrence times $t_{rec}$ are some orders of magnitude greater than the stickiness times $t_{st}$. In fact, for the case 1, the recurrence times are about two orders of magnitude longer than the stickiness times, while for the case 2 the difference is about three orders of magnitude.  Therefore, the orbits stay for a much longer time span inside the chaotic region than inside the sticky region. The black curves in Figs. \ref{distri}b,d are fittings of the distributions derived by eq. (\ref{recft}).
  
We have to point out here that  such extreme sticky phenomena are repeated for all the values of the perturbation parameter $K$, where the diffusion coefficient $D$ presents  maxima (see Fig.1 of \cite{2018PhRvE..97b2215H} ). 
 
  (b) The times of recurrence of individual orbits in the sticky regions are different from each other (even if their initial conditions are all found inside a very small box in the beginning of the calculations). This is a consequence of the fact that the Lyapunov times $t_{L}$ of chaotic orbits are very small, of order 1 or smaller (see Figs. \ref{streth}b,c, from where it appears that in general $LCN$ is larger than 1) thus the deviation of nearby orbits starts very early. Therefore the average value of the diffusion exponent $\mu$ will not deviate considerably from $\mu=1$. A consequence of this diffusion is that the density of the points of an initially sticky orbit, outside the islands of stability tends to become uniform after a long enough time (see \cite{2010IJBC...20.2005C}).

\section{Self-Similarity of Islands}
\begin{figure*}[hbt]
\centering
\includegraphics[scale=0.35]{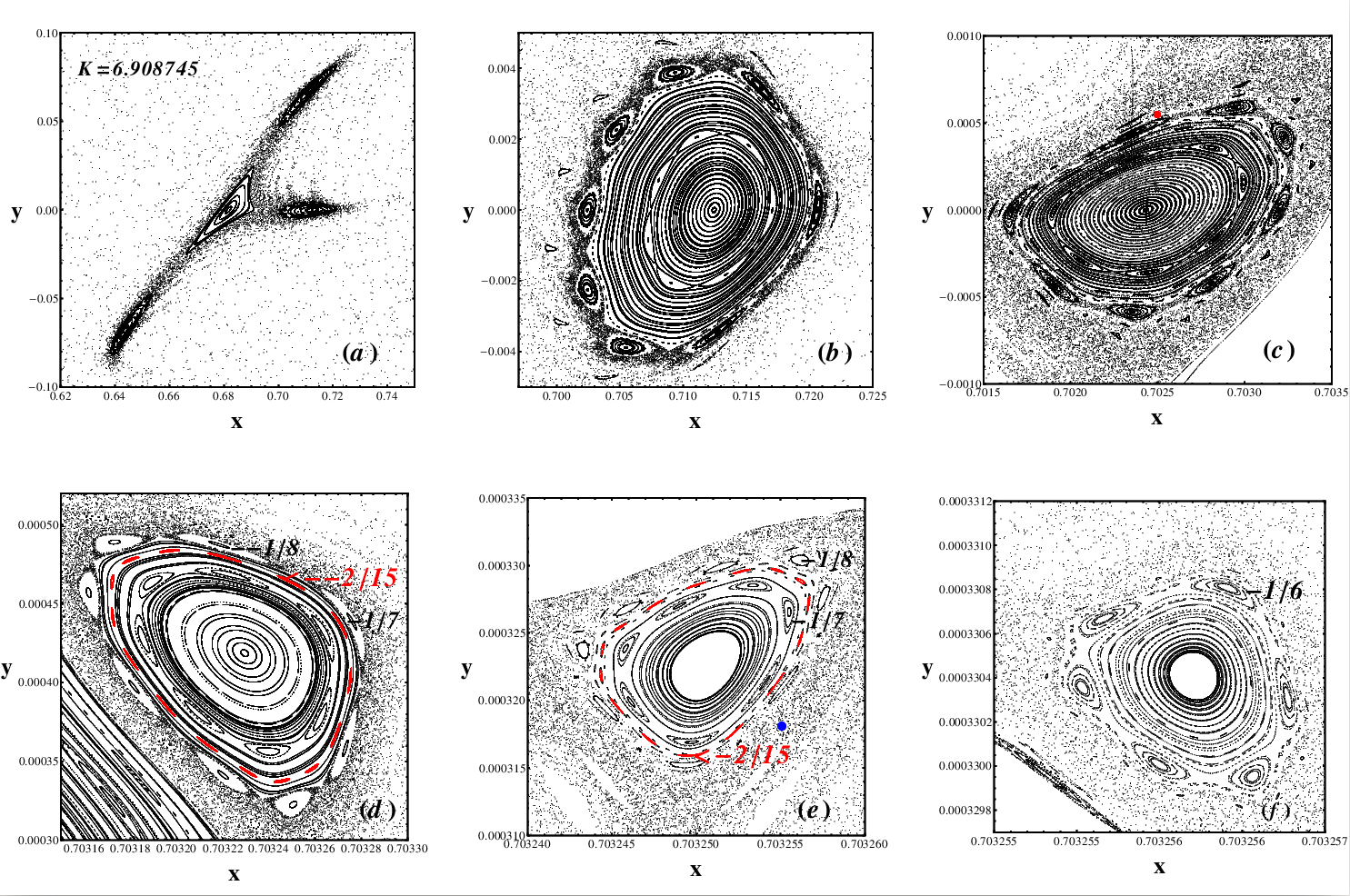}
\caption{ The phase space of the standard map for $K=6.908745$. We plot the hierarchy of the islands of multiplicities 3-8-8-8-8-6 in (a)-(b)-(c)-(d)-(e)-(f) respectively.  } \label{zoom}
\end{figure*}

Stickiness appears around all the islands that exist on the surface of section corresponding to a given value of the perturbation parameter $K$. It is well known that there is a hierarchy of islands that surround periodic orbits generated from the main island of the system by successive bifurcations. As $K$ increases the central stable periodic orbit generates an infinity of higher order periodic orbits until it becomes unstable. A large fraction of these higher order orbits are stable and for larger values of $K$ they generate second order periodic orbits. These orbits generate third order periodic orbits and so on. A particular hierarchy of bifurcations is the period doubling sequence (see e.g. \cite{2002ocda.book.....C}
for a review). 

\begin{figure*}
\centering 
\includegraphics[scale=0.35]{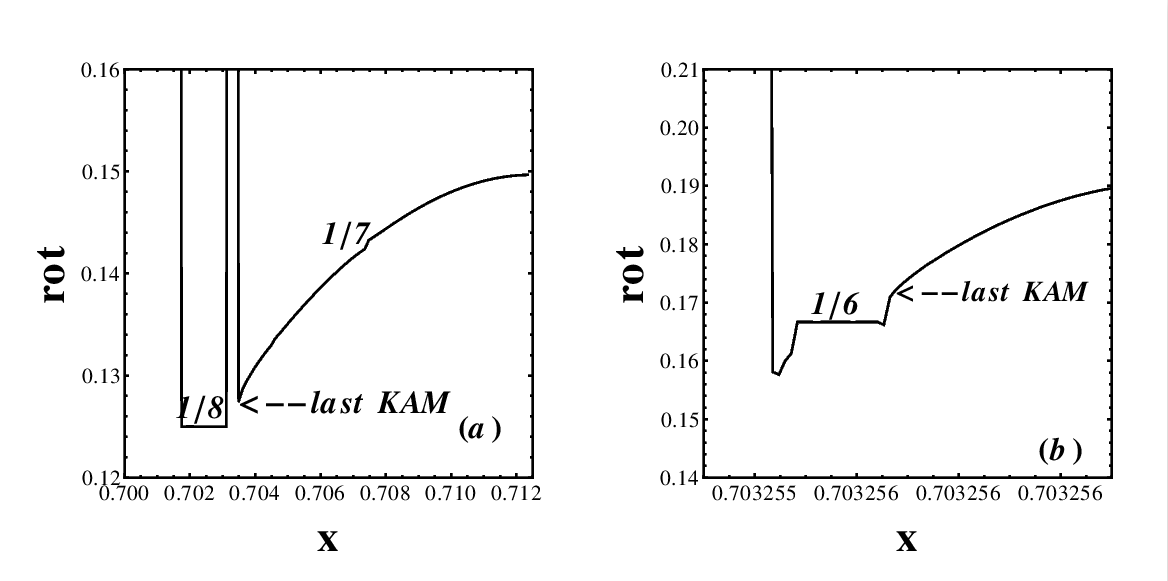}
\caption{ The rotation number as a function of $x$ for the chain of islands of stability corresponding to (a) Fig. \ref{zoom}b, and (b) Fig. \ref{zoom}f. The vertical lines indicate chaotic domains.}\label{rot}
\end{figure*} 
All the stable periodic orbits are surrounded by islands of stability terminating at their respective last KAM curves. The secondary islands, after their generation, are located inside the parent islands, but as $K$ increases they move out of the last KAM curve of the parent island. An example is shown in Fig. 2.107 of \cite{2002ocda.book.....C}, where we see that a change of a parameter $K$ from 4.79 to 4.80 has as a consequence that 5 secondary islands surrounding the central island move from inside the limits of the central island to the chaotic domain outside the last KAM curve of the central island.

This procedure is repeated again and again as $K$ increases and thus we have an infinite hierarchy of islands of successive orders. 

Zaslavsky \cite{zaslavsky1997self}  found a hierarchy of higher order islands of multiplicities 3,8,8,8,... in the case of the standard map with $K=6.908745$. For this value of $K$ there is a stable periodic orbit of multiplicity 2 with 3 islands of stablility around it (Fig.\ref{zoom}a). One of these three islands (on the right) appears in more detail in   Fig. \ref{zoom}b. In this figure we see 8 islands of stability just outside the last KAM curve of the central island. In   Fig. \ref{zoom}c one of these 8 islands is seen in more detail with another 8 islands of stability around it.  The same is found in Figs. \ref{zoom}d,e,f where we show the islands of stability outside the corresponding last KAM curve in more detail at each level. Zaslavsky emphasized the similarity between the 8 islands of stability of successive levels. However, this similarity does not continue in further levels of detail and as we see in Fig. \ref{zoom}f there are 6 islands of stability just outside the last KAM curve and not 8 as in previous levels. 

Furthermore the eight islands in Figs. 10d,c,d,e are not the first islands just outside the corresponding last KAM curves. In fact we find several higher order islands outside the last KAM curves, like for example the islands of stability with rotation number $2/15$ (red islands in Figs. \ref{zoom}d,e). In Fig. \ref{zoom}d these higher order islands are found inside the last KAM curve, while in Fig. \ref{zoom}e they are found outside the last KAM curve.  In Fig. \ref{rot} we plot the rotation number as a function of the x-component across the islands of stability for the cases of Figs. \ref{zoom}b and  \ref{zoom}f respectively. The curve of the rotation number passes through an infinity of rational numbers as we go outwards  along  any given direction from the center. These rational numbers correspond to islands of multiplicity equal to the denominator of the rational number (see \cite{2002ocda.book.....C}).

In particular in Figs. \ref{zoom}(b-e) there are 7 islands of stability inside the last KAM curve of the main island that have rotation number $rot=1/7$. Outside the  last KAM curve of the main island there are 8 islands of stability with rotation number $rot=1/8$. Therefore between them there must be a periodic orbit with rotation number $rot=\frac{1+1}{7+8}=\frac{2}{15}$ (see Figs. \ref{zoom}d,e). Between the islands of multiplicity 15 and the islands of multiplicity 8 there is a periodic orbit with rotation number $rot=\frac{1+2}{8+15}=\frac{3}{23}$ and so on. In fact there is a hierarchy of islands, forming a Farey tree, with multiplicities going all the way to infinity (see \cite{2002ocda.book.....C}).

Therefore there are a lot of higher order islands of stability just outside the last KAM curve.  Neither the 8 nor the 15 islands of stability play the role of $BIC$ (Boundary Island Chain) proposed by Zaslavsky in \cite{zaslavsky1997self} since near the last KAM curves there are islands of multiplicity going to infinity.

\section{Conclusions}

The three characteristic times in the case of the standard map are the Lyapunov time, the Poincar$\acute{e}$ recurrence time and the stickiness time (or escape time). Our main conclusions are as follows:

A) The Lyapunov time $t_L$ is  the inverse of the Lyapunov characteristic number ($LCN$) and it is different in separated chaotic domains, in cases with perturbation parameter $K$ smaller than the critical value $K_c \approx 0.9716...$. The Lyapunov characteristic number $LCN$ for a given perturbation parameter $K<K_c$ is approximately proportional to the square root of the percentage of the area of the corresponding chaotic domain: $LCN=c \sqrt{A}$. On the other hand for $K>K_c$ the value of $LCN$ is related to the corresponding percentage of the available chaotic area $A$ through the relation: $LCN =cA^{4.2}$. For large enough values of $K$ the percentage of the chaotic area $A$ is close to $100\%$ and the $LCN$ is roughly proportional to $\ln K$, i.e. for large $K$ the relation between $LCN$ and $K$ is: $LCN \approx 0.925 \ln K$, while for relatively small $K$ (but $K>K_c$) the relation is: $LCN \approx \ln(0.55+0.46 K)$.

B) The mean Poincar$\acute{e}$ recurrence time $\langle t_r\rangle$ depends on the size $\epsilon$ of the box inside which we take the initial conditions. We have shown that the logarithm of the mean Poincar$\acute{e}$ recurrence times $\langle t_r\rangle$ for all the  chaotic domains (for $K<K_c$ as well as for $K>K_c$) is a linear function of the logarithm of $\epsilon$, with the same slope equal to -2. On the other hand the slope inside the islands of stability is equal to -1. The absolute value of this slope is  equal to the fractal dimension $d_w$,  i.e. $d_w=|slope|$. Moreover, the fractal dimension is equal to : $d_w=2/ \mu$ (see \cite{2005drfd.book.....B}), where $\mu$ is the diffusion exponent. Inside the chaotic domain of the phase space the diffusion exponent converges always to $\mu=1$ and therefore the fractal dimension is $d_w=2$. Inside an accelerator mode island the diffusion exponent is  $\mu=2$ and therefore $d_w=1$. Inside the sticky domain outside an accelerator mode island the diffusion exponent  is $\mu=2$ for some time (and hence $d_w=1$) but for long  enough time it tends to 1, i.e. the diffusion is normal (and hence  $d_w=2$). 

The mean Poincar$\acute{e}$ recurrence time $\langle t_r\rangle$ depends also on the area of the available chaotic domain $A$. In fact the relation between these two quantities for $K<K_c$ as well as for $K>K_c$ is: $<t
_r> \approx A \epsilon ^{-2}$. For large enough $K$, where the percentage of the available area $A$ converges to 1, the mean Poincar$\acute{e}$ recurrence time $\langle t_r\rangle$ is simply proportional to the inverse of the size of the area: $t_r \approx \epsilon^{-2}$. In general  $t_r \approx \epsilon^{-d_w}$, where  $d_w$ is the fractal dimension of the mapping.

For $K<K_c$  the Lyapunov characteristic number $LCN$ is proportional to $\sqrt{A}$, so it follows that: $LCN=g \epsilon \sqrt{t_r}$. As a consequence the Lyapunov time $t_L$ is inversely proportional to the square root of the recurrence time, i.e. for large recurrence time the Lyapunov time is small. However the recurrence time cannot be smaller than a minimum (see \cite{1996CeMDA..63..189C} and \cite{2004CeMDA..88..163C}).  

On the other hand, for $K>K_c$, the relation between $LCN$ and available chaotic area $A$ is found to be: $LCN=cA^{4.2}$ and therefore the relation between $LCN$ and  mean Poincar$\acute{e}$ recurrence time $\langle t_r\rangle$ is: $LCN \propto\langle t_r\rangle^{4.2}$. 

C) The stickiness (or escape) time depends on a number of factors that were discussed in a previous paper of our group \cite{2010IJBC...20.2005C}. The most important factor is the distance from the boundary (last KAM curve) of the island of stability. In fact it was found that as the distance from the island increases the stickiness time decreases almost exponentially. On the other hand very close to this boundary the stickiness time increases considerably and tends to infinity as the distance tends to zero. However these very long stickiness times affect a very small domain close to the boundary of the islands of stability.

D) We studied the distributions of the stickiness times and the Poincar$\acute{e}$ recurrence times and we compared the average values of these times for sets of orbits in small boxes at different distances from the islands of stability.

We found that the Poincar$\acute{e}$ recurrence times are many orders of magnitude larger than the stickiness times. 

E) In the past many authors considered the diffusion exponent $\mu$ of the chaotic part of the phase space, in cases where accelerator mode islands exist, to lie between the values 1 and 2, i.e. they claimed that the overall diffusion exponent for initial conditions in the sticky region around these islands is about 1.5.  
However, we have shown that for initial conditions inside the extreme sticky regions around accelerator mode islands,  if we calculate the diffusion coefficient for long enough time (of the order of $10^{10}$ or more) the diffusion exponent $\mu$ tends always to the value $\mu =1$ and therefore the diffusion is normal. In fact orbits with initial conditions close to the boundary of an accelerator mode island have initially  $\mu=2$ for times corresponding to the stickiness time, but later on when the orbits are detached from the sticky region of the island and for a long interval of time the diffusion exponent remains, close to $\mu=0$. Finally, when these orbits scatter inside the large chaotic sea the diffusion exponent converges to $\mu=1$.  The fact that the orbits come again close to the boundary of the accelerator mode islands infinitely many times, does not affect the average value of $\mu$ because (a) the recurrence times are many orders of magnitude longer than the stickiness times and (b) the recurrence times inside the sticky zone are very different for each initial condition (even if these initial conditions are all found inside a very small box in the beginning of the calculations).  Therefore, the diffusion of these orbits will finally become normal (with $\mu=1$) after long enough computational time. On the other hand, initial conditions in the large  chaotic sea far away from the islands of stability will manifest normal diffusion ($\mu=1$) much earlier.

F) Finally we studied the hierarchy of islands considered by Zaslavsky et al.  in \cite{zaslavsky1997self}, who found a sequence of 3,8,8,8,8,... islands at successive levels (islands around islands). However, we found (a) that this sequence does not continue indefinitely because  at the next level there are 6 islands of stablility around the main island (and not 8) and (b) because the 8 islands do not form a ``boundary island chain'' (called ``BIC'' in \cite{zaslavsky1997self}). In fact closer to the boundaries of the successive sets of 8 islands we found 15 islands (in some levels these islands are inside the last KAM of the main island of stability and in other levels they are outside the last KAM curve). Furthermore, very close to the boundaries of the islands of various levels there are infinite islands of higher and higher multiplicity and therefore one cannot claim that any one of them is ``BIC''. As a consequence there is no self similar sequence of islands around islands, but infinities of islands of various multiplicities.

\begin{acknowledgements}
The authors would like to express their gratitude to Dr. A. C. Tzemos for technical assistance.
This work was partly supported by the Research Committee of the Academy of
Athens, with project number 200/895. 
\end{acknowledgements}   

\bibliographystyle{plain}
\bibliography{characteristic_time}

\end{document}